\documentclass[sigconf,nonacm]{acmart}

\usepackage{listings}
\usepackage{algorithm}
\usepackage{algorithmic}
\usepackage{booktabs}
\usepackage{xcolor}
\usepackage{enumitem}
\usepackage{url}
\usepackage{tabularx}
\usepackage{array}
\usepackage{float}

\definecolor{duckyellow}{RGB}{218,165,32}

\AtBeginDocument{%
}

\renewcommand\footnotetextcopyrightpermission[1]{}
\title{The Garden of Forking Paths: Threading Narrative Archetype as a Semantic Signal Through Gameplay Planning}

\subtitle{%
  \textcolor{duckyellow}{%
    \textbf{Live Demo:} \url{forkgarden.darkmesh.art}%
  }%
}

\author{%
\makebox[0pt][c]{%
\parbox{\textwidth}{%
\raggedright
\textbf{Yunge Wen$^{1,2}$, Chenliang Huang$^{2}$, Hangyu Zhou$^{2}$, Zhuo Zeng$^{2}$, Yuxuan Weng$^{3}$, Timothy Merino$^{2}$, Julian Togelius$^{2}$, Max Kreminski$^{4}$, Sam Earle$^{2}$}\\[2pt]
\large
$^{1}$Massachusetts Institute of Technology \quad
$^{2}$New York University \quad
$^{3}$Nanyang Technological University \quad
$^{4}$Cornell Tech
}}%
}

\begin{document}

\begin{abstract}
Generative models can produce individual game facets, but whole-game generation remains an orchestration problem: narrative, level structure, encounters, objectives, rewards, and visuals must express shared intent. We present \textbf{Forking Garden}, a branching game generation system that uses narrative archetype as a persistent semantic signal across the generation pipeline. Narrative progression is represented as soft Rise/Fall states; candidate plot nodes are generated before structural constraints assemble archetype-conforming graphs. The same state then conditions encounter composition, objectives, rewards, and runtime difficulty adaptation, while a shared symbolic schema preserves narrative entities and gameplay configurations through content instantiation. Across 10 storylines, generated paths exhibit distinguishable archetypal trajectories, narrative threat predicts damage taken by a threat-blind combat agent, and generate-first-constrain-later yields 2.6$\times$ the entity diversity of a hierarchical baseline. A 16-participant study further suggests that propagated Rise/Fall distinctions can remain meaningful during play, while also supporting narrative understanding and creator-oriented interpretation.
\end{abstract}

\begin{teaserfigure}
  \centering
  \includegraphics[width=\textwidth]{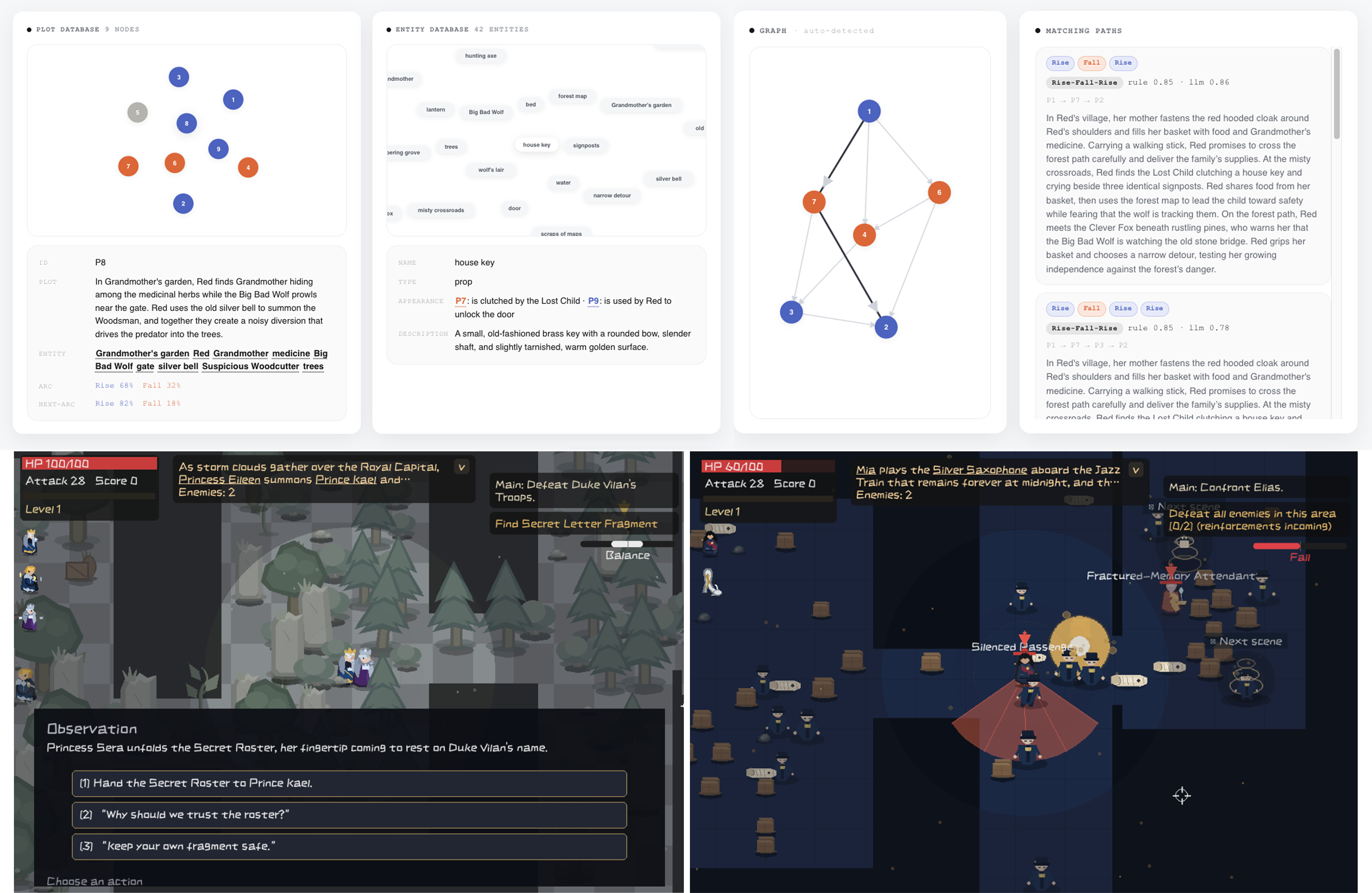}
  \caption{Forking Garden treats narrative archetype as a semantic signal threaded through every step of generation: given a storyline, the system generates independent narrative nodes with extracted entities (top left), assembles them into an archetype-conforming branching graph (top center-left), and instantiates each node with schema-constrained entity art and narrative-grounded content (top right). Every downstream system, from difficulty adjustment to character animation and combat effects, is built on the same symbolic schema, making the entire engine amenable to neuro-symbolic generation. The resulting nodes are playable as dungeon levels, where narrative arc conditions enemy encounters, dialogue, and gameplay objectives (bottom).}
  \label{fig:teaser}
\end{teaserfigure}

\maketitle

\section{Introduction}

As generative AI advances, individual game components become easier to produce. Game designers can now leverage an increasing range of competent foundation models to generate visual content, audio, and 3D assets. LLM-based agentic pipelines can help them generate scripts and control levels by interacting with game engines such as Unity and Unreal. A new question emerges: with so many different game components, how can we combine and integrate them into a complete and coherent gameplay experience~\cite{liapis2019orchestrating}?

Before the LLM era, Liapis et al. had already raised this problem, which they framed as \textit{game facet orchestration}. They described this problem as a spectrum, where most generation approaches can be characterized as bottom-up, top-down, or a mixture of the two. Some approaches use multiple independent generators to produce drafts in a shared workspace, where other generators can access the generated content and transform it into their own inputs for further generation. However, this approach faces a major challenge: visual and audio content can be represented as individual assets and passed between generators, while the more variable semantics of narrative and game rules are difficult to quantify through predefined rules. Another approach is to provide a \textit{composer} with high-level instructions, which are then passed down to subordinate generators. Each subordinate generator interprets these instructions and concretizes them for the next layer. Today, this approach is more feasible because most generative models can take natural language as input and can therefore be coordinated through instructions generated by LLMs.

Structuralist narratology has spent nearly a century describing precisely this kind of shared intent: beneath the enormous variety of stories lies a relatively small set of recurring narrative motifs.  Humans have told stories since prehistoric times, and these stories are abstractions of the world: they describe who did what, where, and why, and are important to how people understand the world and how societies and cultures are formed. When older generations tell stories to children, they pass down these understandings of the world, allowing certain narrative patterns to recur across generations. The structure of “departure--ordeal--return” in the \textit{Odyssey}, for example, established a Western narrative tradition that has persisted for millennia. Through his study of a large corpus of Russian folktales, Propp abstracted dozens of specific recurring narrative functions, such as a hero receiving a magical gift from a helper or being betrayed by someone close to them. These narrative motifs naturally provide semantic signals for symbolic generation because these relationships do not need to be learned from scratch: a motif already conveys what is happening at a particular moment in a story. This is the information a designer needs when deciding what a player should encounter upon entering a game level and how dangerous or fortunate that encounter should be. 

\textbf{Forking Garden} is our attempt to operationalize this idea. Built on top of a symbolic game engine, the system takes a storyline and a protagonist as input and carries the same narrative signal from narrative planning through to the final playable game. The system follows a \textit{generate-first-constrain-later} strategy. It first generates a set of independent candidate narrative nodes, avoiding premature narrative framing that could unnecessarily restrict the space of possible game entities downstream. Each node is assigned soft Rise and Fall scores. A graph-planning process then organizes these nodes into a branching structure, scores root-to-leaf paths, and retains only those that conform to the target narrative archetype. The same Rise/Fall states subsequently guide the generation of game components, including enemy composition, gameplay objectives, and rewards. They also regulate combat pacing in response to player performance without undermining the tension or relief that the current narrative node is intended to convey. Finally, each node is instantiated as a playable level through a schema-constrained transformation, including the corresponding character and item art assets.

We evaluate the system at two levels. First, we conduct a technical evaluation to determine whether the narrative signal is preserved throughout the generation pipeline. Our results show that the signal remains measurable even after leaving the narrative-planning stage. Across ten storylines, accepted paths exhibit overall trajectories consistent with their respective narrative archetypes; narrative threat levels are reflected in the damage sustained by a simulated combat agent that has no direct access to narrative threat information; and the \textit{generate-first-constrain-later} approach achieves a mean entity-diversity ratio of 0.678, compared with 0.261 for a hierarchical baseline. We then conduct a user study to examine how much of these differences ultimately reaches the player. Participants perceived differences between gameplay experiences produced under Rise and Fall conditions, although the strength and interpretation of these differences varied across participants. Some associated these differences with variations in tension or difficulty, while others interpreted them primarily through narrative progression and branching structure.

This paper makes three contributions:
\begin{itemize}[leftmargin=*]
\item We propose an orchestration mechanism built on a persistent narrative signal. Forking Garden represents narrative archetype as soft Rise/Fall state and maintains it as shared state across branching narrative planning, gameplay generation, and runtime adaptation rather than consuming it once at the prompt; structured content instantiation then preserves the resulting narrative and gameplay decisions in executable form.
\item We provide technical evidence that the signal remains reflected in downstream generation. Archetype labels correspond to distinguishable path trajectories, narrative threat is reflected in the damage taken by a combat agent with no access to $T_i$ or archetype information, and generating before constraining preserves substantially more entity diversity than hierarchical refinement.
\item We conduct a 16-participant user study examining whether arc-conditioned differences become perceptible and meaningful during play, complementing the technical evaluation of signal propagation.
\end{itemize}

\section{Related Works}
\subsection{Computational Narratology}

The intuitive observation that stories repeatedly exhibit similar structures has a long history of formal study in narratology. Propp's morphological analysis of Russian folktales identified 31 recurring narrative functions, including villainy,'' departure,'' struggle,'' and return.'' His analysis showed that even when stories differ substantially in their characters and settings, their underlying sequences of functions recur in relatively stable forms~\cite{propp2010morphology}. Campbell proposed a similar regularity at a broader scale. His concept of the monomyth'' describes myths across cultures as sharing a cyclical structure of departure--initiation--return,'' which can be further decomposed into stages such as the call to adventure, trials undertaken with the aid of a mentor, and an eventual return with some transformative gain~\cite{campbell2008hero}. Vogler subsequently adapted this abstract structure into a practical twelve-stage template for screenwriters, translating Campbell's anthropological observations into an operational framework for commercial storytelling~\cite{vogler2007writers}.

In interactive media, prior work has further suggested that different stages of these narrative archetypes can correspond not only to plot events but also to concrete gameplay states, such as resource scarcity near a climax or assistance from a mentor early in the story~\cite{lebowitz2011interactive}. Research centered on emotion and conflict has examined more fine-grained objectives. Hernandez et al. designed experiences in which players progress along designer-specified emotional trajectories~\cite{hernandez2014pace}; Ware et al. formalized dramatic conflict itself as a computable property of a planning-based narrative fabula~\cite{ware2014computational}; and Zook et al. modeled task difficulty through a time-varying ``player skill'' signal that can both be fitted from data and used to drive content generation~\cite{zook2015skill}. Together with Lebowitz's account of the relationship between narrative stages and gameplay, these works demonstrate that emotion, conflict, and difficulty associated with narrative structure can be represented as explicit and testable objectives in interactive systems. However, they generally address these properties individually rather than organizing them under a complete global narrative archetype.

More recently, Reagan et al. provided a fully computational characterization of narrative trajectories. Using sentiment analysis and dimensionality reduction on a large corpus of English-language fiction, they reduced emotional trajectories to combinations of two fundamental directions of change: Rise and Fall~\cite{reagan2016emotional}. Taken together, this line of work shows that narrative archetypes not only recur, but can, at least under Reagan et al.'s formulation, be represented as descriptive patterns that machines can compute. What these approaches do not establish, however, is how such patterns can be used \textit{prescriptively}. They can identify the narrative archetype exhibited by an existing text, but do not specify how a generative system should be constrained by a target archetype so that the content it produces conforms to that archetype.

\subsection{Game Orchestration}

Game generation has long required the coordination of heterogeneous facets, including narrative, levels, mechanics, visual content, and other forms of game content. These facets must not only function individually but also remain mutually consistent. Early systems relied primarily on human intervention to achieve such coordination. Collaborative generation systems, mixed-initiative authoring tools, and generators driven by authorial intent allow creators to continuously inspect, revise, or constrain generated components as a game takes shape~\cite{cook2017angelina1, cook2017angelina2, nelson2017fluidic, kreminski2020germinate, osborn2019generator}. Liapis et al. further formalized this broader problem as game facet orchestration, arguing that coordinating multiple specialized generators is itself a problem distinct from the quality of any individual generator~\cite{liapis2019orchestrating}. In these systems, however, overall coherence is still maintained primarily by the creator or by manually specified constraints; there is no semantic state that independently persists throughout the generation process while continuously carrying authorial intent.

Another line of work has begun to delegate some of this coordination to structured representations. TropeTwist and Story Designer explicitly represent narrative structure to support computational authoring~\cite{alvarez2022tropetwist, alvarez2022storydesigner}, while other systems allow one narrative representation to constrain an adjacent game facet: GB Studio generation maps structured descriptions into executable games~\cite{duplantis2021gbs}, PlotMap generates spatial layouts from relationships specified in a narrative~\cite{wang2024plotmap}, and StoryVerse uses planning over world states to translate abstract narrative actions into character behavior~\cite{wang2024storyverse}. These systems reflect an important shift: different facets are no longer generated entirely independently, and one representation can instead determine another. However, the representation responsible for coordination typically terminates after a single transformation. A narrative may determine a map or a character behavior, but the same semantic state that produced that result does not continue to participate in subsequent generation decisions.

This limitation becomes more pronounced as LLM-based systems substantially reduce the cost of generating heterogeneous content. Modern generation pipelines can delegate text, code, visual assets, animation, and interaction logic to different models or tools. As a result, access to sufficiently capable generators is no longer the primary bottleneck; coordinating their outputs has instead become an increasingly important challenge. Recent LLM-based game-generation systems have likewise begun to address generation at the level of the overall system~\cite{jennings2024gromit, zhang2026rpgagent, xu2026diaryplay, yang2025gptgames}. Yet orchestration in these systems is still typically implemented through repeated prompting, agentic planning, or artifact-specific schemas, rather than through a semantic representation that persists across the entire pipeline. Forking Garden explores precisely this design space: we represent a narrative archetype as an explicit state that persists through successive transformations, allowing it to shape branching structure, encounter composition, gameplay objectives, rewards, runtime adaptation, and ultimately content instantiation, rather than disappearing after it has been consumed by a single generation stage.

\subsection{Neuro-Symbolic Generation}

Neuro-symbolic approaches originally sought to combine the pattern-recognition capabilities of neural networks with the verifiability and compositionality of symbolic representations~\cite{garcez2020neurosymbolic}. Game engines already implement a domain-specific form of such a symbolic world: entities, states, rules, and their relationships are explicitly represented in data structures that determine game behavior. LLMs can generate and manipulate these structured representations, allowing semantically rich outputs to be directly inspected and executed by a game engine without first recovering structure from unconstrained natural language. 

This general framework has been applied in several forms. Symbolically representing the generative workflow allows models to map natural language into editable structures instead of just producing the final artifacts~\cite{Chen_2025_CVPR}. Other approaches incorporate formal constraints into generation~\cite{pmlr-v267-banerjee25a}, or first decompose complex tasks into plans and then use symbolic verifiers to check individual steps before execution~\cite{pmlr-v267-petruzzellis25a,chi2025instructflow}. Therefore, these symbolic structures serve primarily to constrain whether a generated artifact satisfies a particular specification, but once the verification succeeds, the constraint will not be propagated into subsequent stages.

Another line of work uses LLMs to generate multimodal assets. LLMs today have inherent knowledge of structured output formats such as JSON, SVG, and domain-specific languages for motion. The resulting programs can then be verified against geometric or logical properties recovered from the prompt~\cite{ma2025trajectory, tseng2024keyframer, mover2025}. Compared to diffusion-based generative models, these instructed outputs are more explicit and editable. These systems demonstrate that visual content can also be generated in a form readable by symbolic systems, a property on which our asset-generation pipeline relies. Forking Garden employs similar technical mechanisms to propagate narrative archetypes as a shared semantic signal across narrative planning and game asset generation.

\section{Design Goals}

Three design goals follow from the gaps above. The first two concern \textit{narrative planning}, where branching content must be archetype-consistent without collapsing into repetition; the third concerns \textit{gameplay generation}, where that structure has to reach concrete, playable mechanics.

\textbf{G1: Narrative archetype should be an explicit, verifiable semantic signal.} Narrative content should first be generated independently of any structural framing, and only afterward assembled and filtered against a target archetype, so that every root-to-leaf path can be evaluated on its own merits and dramatic tension becomes a property the system can verify.

\textbf{G2: The same signal that constrains narrative structure should also drive gameplay mechanics and stay consistent as it propagates.} The state that determines archetype conformance should be the same state that determines difficulty, objectives, rewards, and pacing, from generation through runtime, so a player's sense of tension can be traced back to their node's narrative arc; and because the structure is a branching graph, this signal must also stay consistent across branches, so a fact established on one path is never treated as established on another.

\textbf{G3: Downstream content should preserve the structured representation established during planning.} Narrative entities and gameplay configurations should remain explicit as they are instantiated into dialogue, visual assets, animation, and executable game objects, so that downstream generation does not discard the structure established upstream. A shared schema makes these outputs directly recognizable and triggerable by the engine without a separate rigging or extraction step.

\begin{figure*}[t]
\centering
\includegraphics[width=\linewidth]{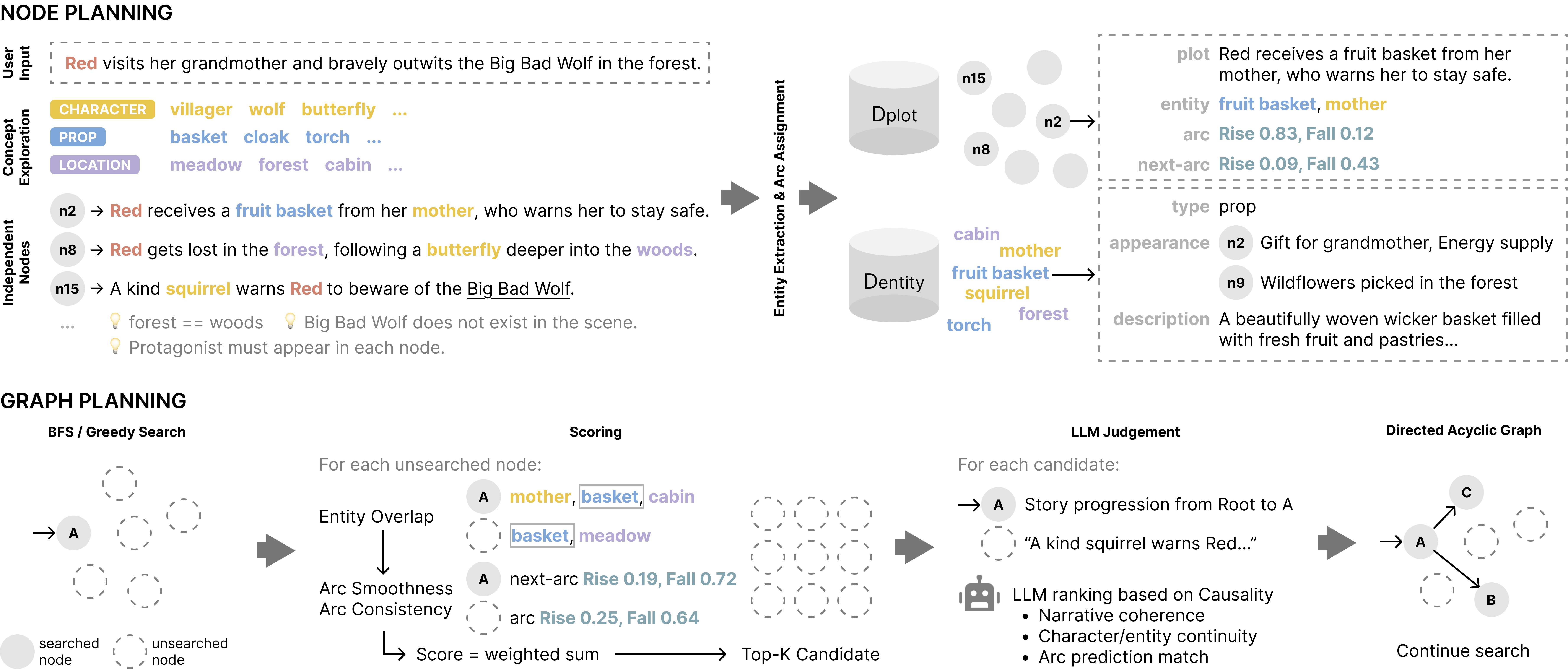}  
  \caption{Overview of the narrative planning pipeline. \textit{Node Planning} (top): given a storyline and protagonist, the system brainstorms candidate characters, props, and locations, then generates independent narrative nodes conditioned on this pool rather than on one another, storing each node's plot content and extracted entities in $D_\text{plot}$ and $D_\text{entity}$ along with soft Rise/Fall arc confidence scores. \textit{Graph Planning} (bottom): starting from the root, candidate successors are scored by entity overlap and arc consistency/smoothness, the top-$k$ candidates are passed to an LLM for causal and narrative-coherence judgment, and the selected edges are added to a growing directed acyclic graph, repeating via breadth-first search until the graph is complete.}
\label{fig:pipeline}
\end{figure*}

\section{Method}

\subsection{Theoretical Background}

In this work, we define a narrative arc as a sequence of discrete beats or discourse segments, while a narrative archetype represents the global story pattern composed of these arcs.

We adopt Reagan et al.'s~\cite{reagan2016emotional} computational framework for our gameplay planning. Reagan et al. applied machine learning techniques including singular value decomposition, hierarchical clustering, and self-organizing maps to analyze sentiment trajectories in a large corpus of English fiction. Their analysis identified two primitive emotional patterns, Rise and Fall, that serve as foundational components across diverse narratives. By composing these primitives into sequences, they identified six narrative archetypes:

\begin{itemize}
\item \textbf{Rags to Riches} (Rise): a continuous ascent from adversity to success
\item \textbf{Tragedy} (Fall): a protagonist's downward spiral toward ruin
\item \textbf{Man in a Hole} (Fall to Rise): overcoming challenges after experiencing setbacks
\item \textbf{Icarus} (Rise to Fall): overconfidence leading to downfall
\item \textbf{Cinderella} (Rise to Fall to Rise): ultimate triumph despite temporary misfortune
\item \textbf{Oedipus} (Fall to Rise to Fall): initial gains followed by catastrophic reversal
\end{itemize}

We adopt this Rise-Fall formulation as the structural constraint mechanism for our narrative planning: its compositional, discrete representation allows a target archetype to be expressed as a sequence of checkable states, well suited to the generate-then-verify approach we take in our graph planning stage.

\subsection{Narrative Planning}

Our narrative planning task takes as input a tuple (Storyline, Protagonist) and produces a branching narrative graph whose nodes can subsequently be instantiated as playable dungeon levels. Each node is associated with soft arc confidence scores $(R,F)\in[0,1]^2$ representing Rise and Fall states. A root-to-leaf path therefore induces a sequence of narrative states that can be evaluated against a target archetype such as Cinderella or Icarus.

We adopt a \textbf{generate-first-constrain-later} paradigm that deliberately separates narrative possibility from narrative structure. Rather than recursively generating each node from its parent, we first generate a diverse pool of mutually independent narrative nodes. Structural and archetypal constraints are imposed only afterward, when these nodes are assembled into a branching graph and complete paths are evaluated against target archetypes. This separation prevents early structural commitments from narrowing the space of downstream narrative possibilities.

All intermediate outputs are represented symbolically. LLMs produce structured descriptions of plots and entities that can be inspected during narrative planning and consumed directly by subsequent gameplay and content-generation stages.

Throughout this section, we use Little Red Riding Hood as a running example.

\subsubsection{Node Planning.}

The first stage expands the input storyline into a pool of independent candidate nodes stored in a plot database $D_{\text{plot}}$. We begin with an LLM-powered brainstorming pass that proposes a broad set of narrative elements, including potential characters (allies, enemies, and neutral figures), objects and tools, locations, and thematic elements such as danger, hope, or betrayal. Candidate plots are then generated from this shared pool together with the original premise, rather than from previously generated plots. Each candidate is a standalone 1--2 sentence narrative event in which the protagonist must appear.

This independence is central to the generate-first-constrain-later design. In a hierarchical generation process, a parent event such as ``Red visits grandmother and outwits the Big Bad Wolf in the forest'' establishes a local framing that naturally biases its descendants toward the grandmother, the Wolf, and the forest. In our approach, no candidate inherits another candidate's framing. The system can therefore explore alternative combinations of characters, objects, locations, and events before committing to a particular narrative structure.

We next extract the entities physically present in each candidate node and store them in an entity database $D_{\text{entity}}$. Each entity is assigned a type (e.g., character or prop) together with its actions or roles in the corresponding plot (e.g., `meets Red'' or `gives a silver whistle''). Physical presence is enforced explicitly: for example, in `Red packs a basket for her grandmother,'' the grandmother is not extracted because she is referenced but does not appear in the scene. Newly extracted entities are matched against existing entries in $D_{\text{entity}}$ so that surface variants such as `the Wolf'' and ``Big Bad Wolf'' resolve to the same entity. Repeated appearances append node-specific information to that shared entry rather than creating duplicate entities.

Each candidate node is then assigned a soft narrative state. We model both its current arc and its predicted next arc as continuous Rise/Fall confidence scores rather than binary labels. The current state combines sentiment analysis with LLM judgment. We first compute a sentiment score $s$ using the GoEmotions classifier, aggregating emotion probabilities as

$$
s = \sum_j w_j p_j,
$$

where $w_j\in[-1,1]$ denotes the polarity weight associated with emotion $j$, and normalize the resulting score to $\hat{s}\in[0,1]$. In parallel, the LLM estimates Rise and Fall confidence scores for the node's current narrative state, as well as a predicted Rise/Fall state for a plausible successor. The current-state estimates are fused with sentiment through

\begin{align}
R_{\text{final}} &= \alpha \hat{s} + (1-\alpha)R_{\text{curr}},\
F_{\text{final}} &= \alpha(1-\hat{s}) + (1-\alpha)F_{\text{curr}},
\end{align}

where $\alpha\in[0,1]$ controls the contribution of the sentiment classifier. A node is Rise-dominant when $R_{\text{final}}>F_{\text{final}}$ and Fall-dominant otherwise, while the continuous scores themselves are retained for subsequent graph construction and gameplay conditioning.

Node planning therefore produces two linked structured databases. $D_{\text{plot}}$ contains $n$ independent candidate nodes, each storing its plot content, associated entities, current Rise/Fall state, and predicted successor state. $D_{\text{entity}}$ contains the unique entities appearing across these candidates and links each entity back to the nodes in which it is physically present. At this point, no global narrative structure or archetype has yet been imposed: the output is intentionally a pool of possibilities rather than a story graph.

\begin{algorithm}[t]
\caption{Arc-Guided Edge Scoring and Selection}
\label{alg:edge_scoring}
\begin{algorithmic}
\REQUIRE Current node $v_c$, candidate nodes ${v_1,\ldots,v_m}$, threshold $\tau$, pool size $N$, branch count $k$, weights $w_e,w_s,w_a$
\ENSURE Selected successor nodes ($k$ of them)
\FOR{each candidate node $v_n$}
\STATE \COMMENT{Entity Overlap Score}
\STATE $s_e \gets |\mathcal{E}_c \cap \mathcal{E}_n| / \max(|\mathcal{E}_c|,|\mathcal{E}_n|)$

```
\STATE \COMMENT{Arc Consistency Score}
\STATE $s_a \gets 1-\frac{1}{2}(|R_{c,\text{next}}-R_{n,\text{curr}}|+|F_{c,\text{next}}-F_{n,\text{curr}}|)$

\STATE \COMMENT{Arc Smoothness Score}
\STATE $\Delta \gets |(R_{c,\text{curr}}-F_{c,\text{curr}})-(R_{n,\text{curr}}-F_{n,\text{curr}})|$
\STATE $s_s \gets \max(0,1-\Delta/\tau)$

\STATE \COMMENT{Weighted Sum}
\STATE $S(v_c,v_n) \gets w_e s_e+w_s s_s+w_a s_a$
```

\ENDFOR
\STATE $\text{candidates} \gets \text{top-}N\text{ nodes by }S(v_c,v_n)$
\STATE $\text{successors} \gets \text{LLM}(v_c,\text{candidates},k)$
\STATE \COMMENT{Select causally plausible successors}
\RETURN successors
\end{algorithmic}
\end{algorithm}

\subsubsection{Graph Planning.}

Graph planning imposes structure on the independent candidate pool. We deliberately separate this process into two levels of constraint. First, \textit{local edge construction} determines which pairs of nodes can plausibly follow one another. Second, \textit{global path filtering} determines whether complete root-to-leaf trajectories conform to a target narrative archetype. This separation allows local narrative coherence to guide graph construction without prematurely forcing individual generation decisions into a global archetypal template.

For each node being expanded, every candidate node that would not introduce a cycle is considered as a potential successor. Candidate edges are ranked using a hybrid scoring function that captures three forms of local continuity (Algorithm~\ref{alg:edge_scoring}). The entity overlap score $s_e$ measures continuity in characters and props between the two nodes. The arc consistency score $s_a$ measures agreement between the current node's predicted successor state and the candidate node's current Rise/Fall state. The arc smoothness score $s_s$ penalizes transitions whose net-rise difference exceeds threshold $\tau$. These components are combined as

$$
S(v_c,v_n)=w_e s_e+w_s s_s+w_a s_a.
$$

The top-$N$ candidates under this score ($N=10$ in our implementation) are passed to an LLM together with their plot content and arc states. At this stage, the LLM evaluates \textit{local causal and narrative plausibility}: whether the candidate can reasonably follow the current event given the characters, objects, actions, and narrative transition involved. Importantly, the LLM does not enforce the target global archetype during edge selection. Archetype conformance is evaluated only after complete paths have been constructed.

Starting from the root, we repeat this procedure in breadth-first order and select up to $k$ successors for each expanded node. A node may be reused as a successor from multiple parents provided that doing so introduces no ancestor cycle, producing a directed acyclic graph rather than a strict tree. The resulting DAG captures multiple locally coherent ways of connecting the independently generated narrative events while preserving shared nodes where different branches can plausibly converge.

Only after this graph has been constructed do we impose the global archetype constraint. We enumerate root-to-leaf paths using depth-first search and evaluate each complete trajectory against the target archetype. For a target such as Rise$\rightarrow$Fall$\rightarrow$Rise, a path is partitioned into corresponding segments and evaluated using each node's net rise

$$
\Delta_i=R_i-F_i.
$$

For a Rise segment, the conformance score combines whether $\Delta_i$ tends to increase across the segment with whether its values are predominantly positive; a Fall segment analogously rewards decreasing trajectories and predominantly negative values. This produces a symbolic path-level conformance score that measures whether the sequence of soft narrative states realizes the requested archetypal shape.

The highest-scoring paths are subsequently passed to an LLM for semantic validation. Given the path's plot sequence and net-rise trajectory, the LLM evaluates whether the resulting progression is narratively coherent and whether the computed archetypal transition is expressed naturally by the events themselves. Paths whose validation confidence exceeds threshold $\theta$ are retained, and the final graph is the subgraph induced by these accepted paths.

We additionally support an auto-detect mode in which the same path-level procedure scores each root-to-leaf trajectory against all supported archetypes and assigns the best-fitting one. This mode allows the system to retain coherent trajectories when a finite candidate pool does not contain enough compatible nodes to realize a specifically requested archetype.

The resulting graph completes the generate-first-constrain-later process: node planning first preserves a broad space of independent narrative possibilities; local graph construction then establishes causal connectivity without imposing a global story shape; and only complete trajectories are finally constrained and verified against narrative archetypes. The Rise/Fall state retained on the accepted nodes becomes the persistent semantic representation passed to the subsequent gameplay-generation stage.

\begin{figure*}
    \centering
    \includegraphics[width=1\linewidth]{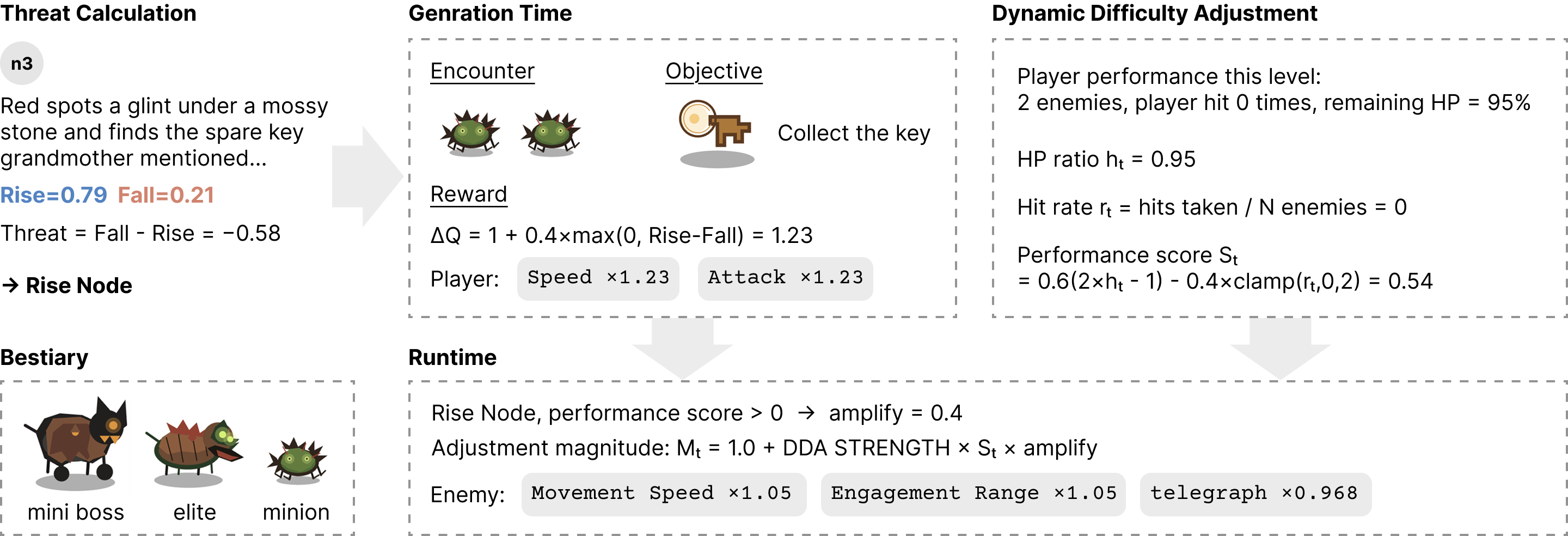}
    \caption{Generation-time and runtime arc conditioning for a single Rise-dominant node. A node's Rise/Fall confidence scores yield a narrative threat $T_i = F_i - R_i$, which conditions encounter composition, objective type, and reward scaling at generation time. The same node's runtime performance score $S_t$, computed from the player's remaining health and hit rate, combines with $T_i$ to determine the dynamic difficulty adjustment magnitude $M_t$, which asymmetrically scales enemy movement speed, engagement range, and attack telegraph time depending on the node's narrative arc.}
    \label{fig:arc-conditioning}
\end{figure*}

\subsection{Gameplay Generation}

After constructing the arc-conditioned branching graph, Forking Garden transforms each narrative node into a playable dungeon level. Game generation preserves the Rise/Fall representation introduced during gameplay planning, allowing narrative progression to influence not only the structure of the branching graph but also the player's moment-to-moment gameplay experience. We separate this process into two stages: \textit{generation-time arc conditioning}, which translates the narrative state of each node into gameplay mechanics and level configurations, and \textit{runtime arc-aware adaptation}, which dynamically adjusts gameplay according to player performance while preserving the intended narrative function of the node.

\subsubsection{Generation-Time Arc Conditioning.} For each narrative node $v_i$, the game generator receives its narrative content, associated entities, graph connectivity, and arc state $A_i=(R_i,F_i)$. We define the narrative threat of the node as $T_i = F_i-R_i$.

The generator first instantiates the entities associated with the narrative node as gameplay objects, including enemies, NPCs, items, and environmental elements, and instantiates outgoing graph edges as portals connecting the corresponding dungeon levels. The arc state then conditions three further gameplay dimensions.

\textit{Encounter composition.} Each tier of enemy carries a fixed, static stat multiplier independent of $T_i$; what $T_i$ actually governs is which tiers, and how many enemies, a level draws, supplementing the node's own entities with enemies from a shared, graph-wide bestiary when needed, with elite enemies appearing only past a threshold and a miniboss appearing with threat-scaled probability. For instance, a Rise node where Red picks up an item draws only a single minion enemy, while a Fall node where Red encounters the Wolf draws the Wolf as the native antagonist plus two elite enemies. This threshold-and-sampling relationship, the mechanism through which a single scalar reshapes an entire encounter, is what ultimately produces the threat-to-damage relationship we measure independently with a threat-blind simulated agent in Figure~\ref{fig:threat-danger} (Section~5.2).

\textit{Objectives.} The same value governs which category of objective a node receives: Fall-dominant nodes are assigned combat-oriented objectives such as defeating all enemies or surviving sustained pressure, while Rise-dominant nodes are assigned collection, NPC interaction, or exploration objectives, and climactic nodes containing the primary antagonist always receive an antagonist-defeat objective. A node's outgoing portals do not open until its objective is met.

\textit{Rewards.} $T_i$ also governs the magnitude of item effects granted for that node, permanent attack-power gains and temporary movement-speed boosts. Rise-dominant nodes scale these effects according to
$$
Q_i = Q_0\left(1+\lambda\max(0,R_i-F_i)\right),
$$
where $Q_0$ denotes the base effect magnitude and $\lambda$ is a scaling constant tuned empirically through playtesting; Fall-dominant nodes leave them at their base value $Q_0$. For instance, a Rise node where Red finds a hidden loaf of bread grants a stronger speed and attack boost, mechanically strengthening the player for the remainder of the level.

Across all three dimensions, the same single scalar, $T_i$, is what varies; what a designer or player observes as "this node feels tense" or "this node feels like a breather" is a direct, traceable consequence of that one number, not a separate authored decision for encounters, objectives, and rewards. The resulting level configuration is stored alongside the original Rise/Fall state, so this dependency remains inspectable after generation rather than only implicit in the output.

\subsubsection{Runtime Dynamic Difficulty Adjustment.} Generation-time conditioning fixes the intended gameplay characteristics of a node, but a fixed difficulty produces substantially different experiences across players of different skill. We therefore adapt gameplay at runtime according to observed performance, estimated from remaining health and normalized hits taken and smoothed over time into a bounded performance score

$$
S_t = c_1(2h_t-1) - c_2 \times \text{clamp}(r_t, 0, 2),
$$

where $h_t$ denotes the player's remaining health ratio, $r_t$ denotes the number of hits taken normalized by encounter size, and $c_1, c_2$ are fixed weighting constants, both computed over the level just completed.

Unlike conventional dynamic difficulty adjustment, in which the adjustment depends only on $S_t$, Forking Garden makes the magnitude of adjustment a function of both player performance and the node's narrative threat,

$$
M_t = 1 + \kappa \cdot S_t \cdot \alpha(T_i, S_t),
$$

where $\kappa$ is a fixed scaling constant and $\alpha(T_i,S_t)$ applies the adjustment asymmetrically with respect to $T_i$: in Fall-dominant nodes ($T_i\geq0$), strong performance ($S_t>0$) sets $\alpha=1$, permitting pressure to increase substantially, while poor performance sets $\alpha=0.5$, earning only limited easing, so the intended tension of the Fall is not erased; in Rise-dominant nodes ($T_i<0$), the relationship inverts, with struggling players ($S_t<0$) receiving $\alpha=1$ and substantial easing, while strong players receive only $\alpha=0.4$ and a small additional increase, so the node keeps functioning as a breather even for a skilled player. A single global DDA rule would treat these two nodes identically at the same performance level; conditioning $M_t$ on $T_i$ through $\alpha$ is what keeps a Fall tense and a Rise restorative regardless of who is playing.

For instance, in a Fall-dominant node culminating in a miniboss encounter, a skilled player who clears it with little damage taken faces a substantially harder encounter on their next attempt; in a Rise-dominant node such as recovering a lost item, a skilled player breezing through receives only a small uptick in difficulty, preserving its function as a moment of respite.

This adjustment is scoped to combat pacing (enemy movement speed, engagement range, reaction timing, concurrent encounter pressure) instead of the base health and damage values fixed at generation time. Rise/Fall therefore acts as a shared control representation across both stages: at generation time it decides what kind of experience a node should be, and at runtime it decides how far that experience is allowed to bend around an individual player, without ever letting the bend erase the intended shape.

\begin{figure*}
    \centering
    \includegraphics[width=1\linewidth]{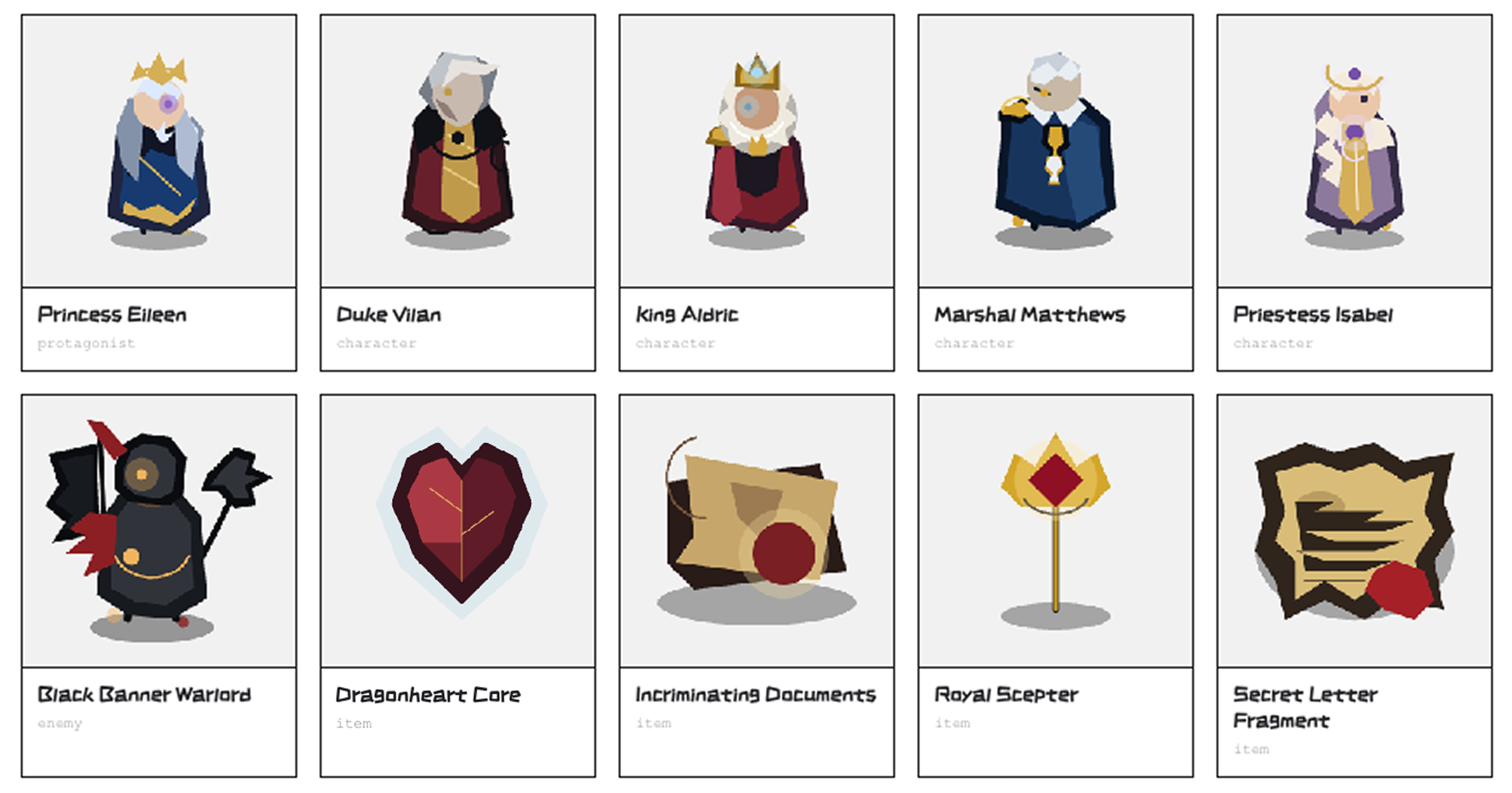}
    \caption{Examples of schema-constrained entities generated by Forking Garden, including the protagonist, supporting characters, an enemy, and narrative items. Each entity is represented as structured vector art that can be instantiated directly as an executable game object while preserving its narrative identity and role.}
    \label{fig:placeholder}
\end{figure*}

\subsection{Content Instantiation}

Content instantiation preserves the entities, continuity constraints, and gameplay configurations established upstream in structured forms that the game engine can directly execute. Rise/Fall governs narrative and gameplay progression; the shared symbolic schema ensures that these decisions are preserved when converted into dialogue, visual assets, animation, and runtime objects.

\subsubsection{Narrative Consistency}

Because Forking Garden's output is a branching graph rather than a single linear story, the same entity or plot thread can carry a different history depending on the path taken to reach a node. Forking Garden therefore performs a narrative grounding pass over the filtered DAG that computes, for every node, which facts are \textit{guaranteed} (true on every path reaching that node) and which are \textit{conditional} (true on only some paths).

This grounding feeds two global passes. A single LLM call fixes each character's identity for the entire cast, keeping a character's voice and self-presentation consistent across branches. A second LLM call performs continuity planning over the full graph, assigning each node the facts it may inherit from earlier nodes and those it establishes for later ones. Where the grounding pass identifies ambiguity, the system prevents facts established on one path from being treated as established on another. Per-node generation then conditions on these fixed identities and continuity bindings.

\subsubsection{Schema-Constrained Art Generation}

Schema-constrained art generation preserves entity identity and engine-level structure while converting planned entities into executable visual objects. Rather than generating raster assets that require separate rigging or sprite extraction, Forking Garden generates visual assets as code. An LLM produces structured, schema-validated vector primitives and animation bindings that a generic rendering executor interprets directly at runtime.

Character and item entities are generated through a two-stage schema. The first stage fixes a silhouette category, color-palette slots, and entity-specific visual features. This specification is cached per entity so that an entity appearing across multiple nodes is drawn consistently. The second stage takes this specification as fixed input and composes the actual vector layers. Generated layers are validated against the rendering schema, with invalid outputs triggering a repair pass.

Because each layer remains an explicit, addressable object, individual components such as limbs, clothing, or weapons can be animated, recolored, or triggered independently at runtime. Scenery uses a lighter-weight process in which a shared library of decor archetypes is reused across locations and tinted according to each location's color scheme.

Appendix~\ref{sec:instantiation-example} provides a concrete example of this process using the Flower Girl character.

\begin{table*}[t]
\centering
\caption{Initial storylines used in the technical evaluation corpus.}
\label{tab:eval-storylines}

\begin{tabularx}{\textwidth}{
    >{\raggedright\arraybackslash}p{0.15\textwidth}
    >{\raggedright\arraybackslash}p{0.15\textwidth}
    >{\raggedright\arraybackslash}p{0.15\textwidth}
    >{\raggedright\arraybackslash}X
}
\toprule
Title & Setting / Period & Genre & Initial Storyline \\
\midrule

Fog Harbor Detective
& Late 19th-century Britain
& Mystery / Detective
& A young detective investigates disappearances in a foggy harbor city and discovers their connection to a passenger ship that sank ten years earlier. \\

Moon Mechanic
& Lunar colony / Future
& Hard Sci-Fi / Survival
& A mechanic apprentice must recover a lost component to repair a failing oxygen system while confronting an old robot and conflicts among factions within the colony. \\

Paper Lantern Market
& Supernatural East Asian setting
& Fantasy / Mystery
& A girl enters a supernatural market that appears only on rainy nights and must trade with spirits and solve puzzles to recover her younger brother's stolen name. \\

Apocalypse Seed Vault
& Post-apocalyptic United States
& Survival / Drama
& A botanist escorts the last batch of fertile seeds to a mountain shelter while facing resource shortages, raiders, and betrayal within the group. \\

Midnight Jazz Train
& Timeless supernatural train
& Surreal / Noir Mystery
& A saxophonist trapped aboard a train that runs forever at midnight discovers that each performance reveals part of a passenger's hidden past. \\

Dragonbone Capital
& Nordic-inspired medieval kingdom
& Epic Fantasy
& A young knight discovers that ancient dragon bones beneath the capital are awakening while the royal family, church, and resistance compete for their power. \\

Candy Factory Escape
& Modern Britain / Magical factory
& Fantasy / Comedy
& A boy trapped in a magical candy factory whose rooms continuously rearrange themselves must cooperate with its strange inhabitants to escape. \\

Deep-Sea Echo Station
& Near-future deep-sea station
& Horror / Psychological Mystery
& A researcher wakes alone in a powerless underwater station and hears distress recordings of herself from three hours earlier while searching for her missing crew. \\

The Last Letter
& Contemporary Chinese old city
& Realism / Emotional Narrative
& Before an old neighborhood is demolished, a postal worker searches for the recipient of a letter left undelivered for twenty years. \\

Time-Loop Sports Day
& Contemporary Japanese high school
& School / Light Sci-Fi
& A student trapped in a repeating school sports day must help her classmates resolve unfinished matters and uncover the cause of the time loop. \\

\bottomrule
\end{tabularx}
\end{table*}

\section{Technical Evaluation}

\subsection{Evaluation Setup}

We assemble an evaluation corpus of $N=10$ initial storylines spanning ten distinct genres and settings: period mystery, hard science fiction, East Asian fantasy, post-apocalyptic survival, surreal noir, epic fantasy, comedic adventure, deep-sea horror, social realism, and school-life time-loop. The initial storylines were generated by an LLM and manually reviewed before being used in the evaluation. Each storyline consists of an initial premise and a protagonist, as summarized in Table~\ref{tab:eval-storylines}.

For each initial storyline, the system first generates 12 candidate plot nodes from its premise and protagonist. Each candidate plot represents a possible narrative event rather than a complete Rise/Fall trajectory. The system assigns Rise and Fall scores to these plot nodes and organizes them into a branching narrative graph. It then evaluates root-to-leaf paths through the graph against the six supported narrative archetypes and retains a target of five archetype-consistent paths for each storyline. The archetype label therefore describes the Rise/Fall trajectory of an accepted path, while the individual candidate plots serve as the nodes from which these paths are constructed. We run each storyline through the complete pipeline, from plot generation and graph planning through content instantiation.

\textit{Archetype distribution.} We tally the auto-detected archetype label of every accepted path across the corpus, giving the corpus-level frequency of each of the six archetypes our system supports.

\textit{Threat vs.\ realized encounter danger.} Reading a danger score directly off the same threshold rule that determines encounter composition (Section~4.3.1) would only confirm that our implementation matches its own specification. We instead measure a downstream, non-deterministic consequence of that rule: for every generated node across all ten games, we instantiate its enemies in our game engine and run a scripted, threat-blind combat agent through the encounter. The agent selects and engages the nearest living enemy using the engine's own hit-detection and damage-resolution code, with no access to $T_i$, tier labels, or archetype information. We record the damage the agent takes before it either clears the encounter or dies, averaged over $n{=}12$ runs per node with randomized enemy placement and a small per-frame chance of a missed decision, so that repeated runs of a node explore different concrete encounters rather than replaying an identical simulation. Because all ten storylines draw from a shared pool of behavior types and tier parameters, many nodes across different storylines resolve to the same enemy composition and therefore near-identical realized damage; we treat node-level observations as clustered by storyline and test the threat--danger relationship with a permutation test that reshuffles $T_i$ only within each storyline, rather than assuming ninety independent observations.

\textit{Net rise by archetype.} For every accepted path, we walk its nodes in order and record each node's net rise, $R_i - F_i$, then group these per-path sequences by the path's archetype label and linearly resample each to a common grid so that paths of different lengths can be overlaid and averaged.

\textit{Entity diversity.} For each storyline, we compute the ratio of unique entities to total entity mentions across its candidate plot pool, and compare this ratio against a hierarchical baseline that conditions each candidate plot on the previous plot's fixed framing, using identical entity extraction and disambiguation for both.

\subsection{Results}

\begin{figure}[t]
\centering
\includegraphics[width=\linewidth]{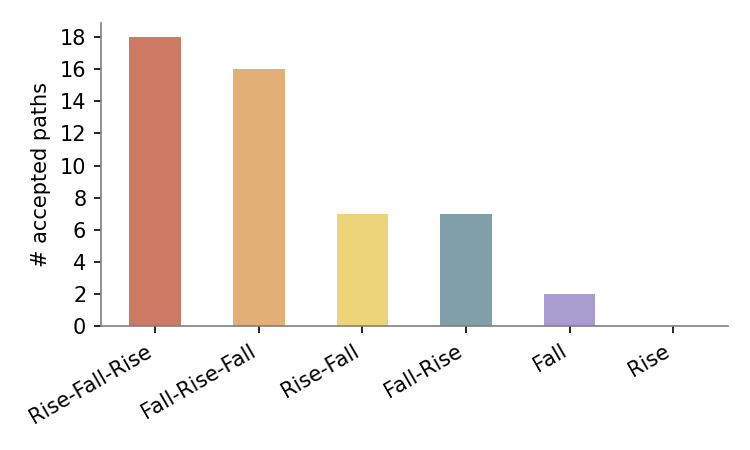}
\caption{Archetype distribution across all accepted paths in the 10-storyline corpus. Three-act patterns (Rise-Fall-Rise and Fall-Rise-Fall) dominate, together accounting for 34 of the 50 accepted paths.}
\Description{Bar chart showing the count of accepted paths per archetype label, dominated by Rise-Fall-Rise and Fall-Rise-Fall.}
\label{fig:archetype-dist}
\end{figure}
\begin{figure}[t]
\centering
\includegraphics[width=\linewidth]{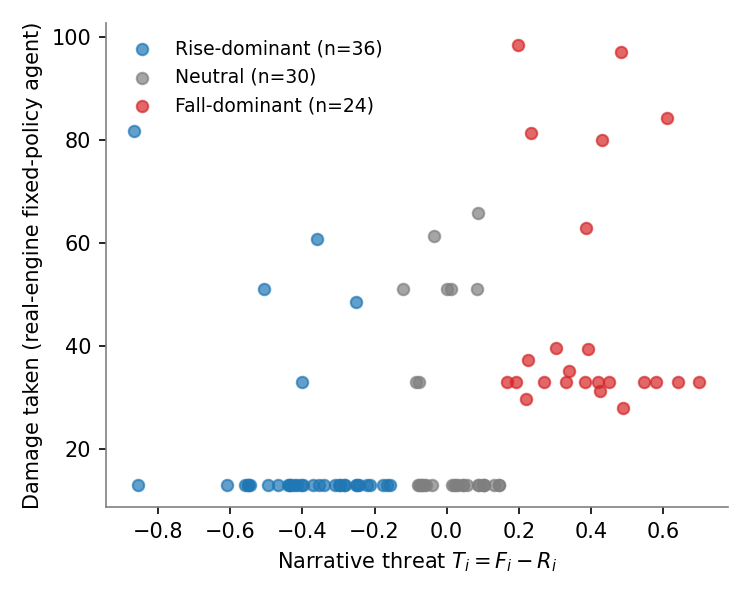}
\caption{Every generated node across all 10 games, plotted by narrative threat $T_i$ against damage taken by a threat-blind fixed-policy combat agent (mean of 12 simulated runs per node with randomized encounter layout), colored by Rise-dominant / Neutral / Fall-dominant. Damage taken rises with threat category (Spearman's $\rho=0.494$; cluster-aware permutation test shuffling $T_i$ within storyline, $p=0.0005$). Horizontal banding reflects the small, discrete set of enemy tiers and behavior types shared across storylines, not an artifact of the agent or the visualization.}
\Description{Scatter plot of narrative threat against damage taken by a threat-blind simulated combat agent, with points colored by whether the node is Rise-dominant, Neutral, or Fall-dominant.}
\label{fig:threat-danger}
\end{figure}
\begin{figure*}[t]
\centering
\includegraphics[width=\linewidth]{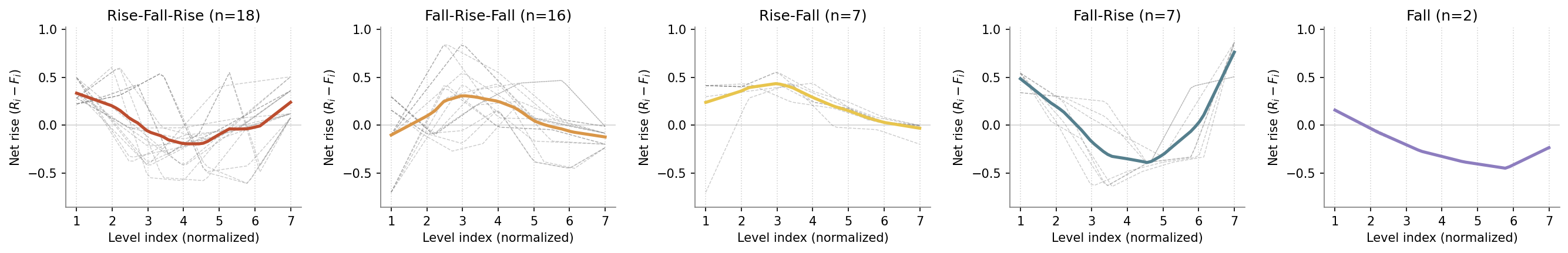}
\caption{Net rise ($R_i-F_i$) along each accepted path, grouped by archetype. Thin lines are individual paths; the bold line is the per-archetype mean. Each archetype's mean curve broadly tracks the shape implied by its label, though individual paths vary considerably around that mean.}
\Description{Five small multiples, one per non-empty archetype, each showing individual path net-rise curves in light gray and a bold colored mean curve.}
\label{fig:net-rise}
\end{figure*}
\begin{figure}[t]
\centering
\includegraphics[width=\linewidth]{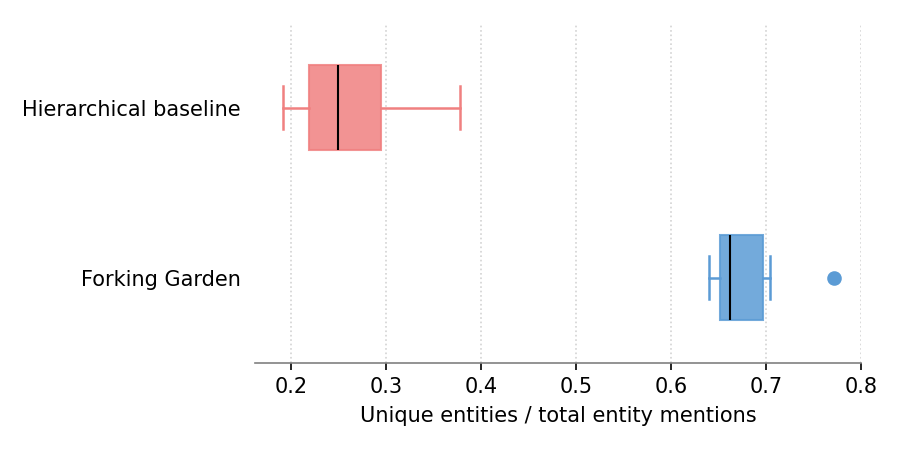}
\caption{Entity diversity ratio (unique entities / total entity mentions) across the same 10 storylines, comparing Forking Garden's generate-first-constrain-later approach against a hierarchical baseline. Forking Garden achieves a mean ratio of 0.678 versus 0.261 for the baseline, with no overlap between the two ranges.}
\Description{Horizontal box plots comparing entity diversity ratio between Forking Garden and a hierarchical baseline, with Forking Garden's distribution entirely above the baseline's.}
\label{fig:entity-diversity}
\end{figure}

The accepted-path archetype distribution is dominated by three-act patterns: Rise-Fall-Rise and Fall-Rise-Fall together account for 34 of the 50 accepted paths across the corpus, consistent with our corpus's use of 12-plot storylines long enough to support a turn and a second turn. Notably, the single-arc Rise archetype was never accepted as the best fit for any path in the corpus, while Fall was accepted only twice; two-act and three-act patterns dominate throughout (Figure~\ref{fig:archetype-dist}).

Damage taken by a threat-blind combat agent increases with narrative threat (Spearman's $\rho=0.494$; cluster-aware permutation test shuffling $T_i$ within storyline, $p=0.0005$, $n_{\text{perm}}{=}2000$): Fall-dominant nodes expose the agent to higher and more variable damage than Rise-dominant nodes, with Neutral nodes intermediate (Figure~\ref{fig:threat-danger}). Because the agent has no access to $T_i$, tier, or archetype information, this relationship cannot be attributed to the agent reading off the rule that generated the encounter; it reflects that narrative threat, propagated only through its effect on which enemies a node contains, produces a measurable difference in realized combat outcome. As in Section~4.3.1, our enemy population is drawn from a small, discrete set of behavior types and tiers shared across all ten storylines rather than a continuously tunable difficulty parameter, so many nodes across different storylines resolve to identical enemy compositions and, consequently, near-identical realized damage; this produces the horizontal banding visible in Figure~\ref{fig:threat-danger}, which we treat as a property of the underlying design space rather than an artifact of the agent or the visualization.

The mean net-rise curve for each archetype broadly tracks the shape implied by its label (e.g., Rise-Fall-Rise paths rise, dip, and recover), though individual accepted paths vary considerably around that mean, particularly in the middle segment of longer paths (Figure~\ref{fig:net-rise}). This indicates that archetype labels assigned during filtering are not merely nominal: the aggregate trajectory of paths sharing a label is visually distinguishable from paths under a different label, even though no single path perfectly reproduces the idealized archetype shape.

Forking Garden's generate-first-constrain-later approach achieves a mean entity diversity ratio of 0.678 (range 0.640–0.771, $n{=}10$), compared to 0.261 (range 0.191–0.378, $n{=}10$) for a hierarchical baseline that conditions each candidate plot on the previous plot's fixed framing, a 2.6$\times$ difference with no overlap between the two ranges (Figure~\ref{fig:entity-diversity}). This confirms that generating candidate content independently of any parent framing, before structural constraints are imposed, substantially reduces the tendency to recombine a small, repeated cast of entities that hierarchical, parent-conditioned generation is prone to.

\section{User Study}
\subsection{Study Design}

\subsubsection{Participants}

We recruited 16 participants through email invitations and messaging-group posts in online gaming communities and university communities, including 10 participants from game-development-related academic programs. Participants ranged in age from 19 to 28, with 6 identifying as male and 10 as female. Participants reported playing a range of game genres, including action, strategy, MOBA, puzzle, narrative, and independent games. Five participants reported prior experience in game design or development. Among them, three had worked on complete independent game projects and had experience across multiple stages of the development process, from game design and planning to implementation. Eight participants had previously used generative AI for game-related creation. Four had used it to assist with programming or game logic, while six had used it to generate or support visual assets, narrative content, or level design. All participants provided informed consent before participating in the study.

\subsubsection{Procedure}

Participants were first given a brief overview of the system's intended purpose and the available means of interaction. Each participant experienced three complete playthroughs: two games pre-generated and stored in the system's community garden, and one game generated live from a storyline of the participant's own choosing.

For the community-garden games, participants were invited to preview the generated storyline, branching graph, and entity art before playing. For their own generated game, participants were encouraged to observe the generation process as the system expanded their storyline into an increasingly detailed plan, entity pool, and branching graph. They were asked to assess the generated content by thinking aloud during this process. Participants then observed the resulting entity art and gameplay assets and examined how the generated content reflected the story they had provided. After each of the three games, participants played the resulting dungeon levels to completion or until they chose to stop.

After using Forking Garden, participants completed a semi-structured exit interview organized around five themes:

\begin{itemize}
\item whether a level's tension or ease tracked the story's narrative archetype
\item whether difficulty appeared to adapt to their own performance
\item whether character, item, and environment art felt visually coherent
\item whether they noticed any contradictory or incoherent dialogue or plot content across branches
\item whether they would want the system as a tool for their own creative use
\end{itemize}

Each study session lasted a maximum of 30 minutes. The user study design, including calls for participation, study structure, participant workflow, and intake and exit interview questions, was approved in accordance with our institution's ethical review requirements prior to conducting the study.

\subsubsection{Data Collection and Analysis}

In addition to the intake survey, the researcher present during each session took notes on participants' interactions with and commentary on the system. Each participant consented to audio recording of the exit interview, and the recordings were transcribed for analysis.

Following guidelines for qualitative HCI practice, the authors collectively reviewed the study notes and interview transcripts and iteratively identified recurring topics and behavioral patterns pertinent to our research questions. We first coded whether participants reported perceiving differences associated with narrative progression or difficulty. We then examined how participants described their branch choices, what information they attended to when making those choices, and what aspects of gameplay they used to explain perceived differences. This second pass distinguished semantic cues encountered during decision making from cues inferred through gameplay experience.

Because we did not retain participant-level runtime logs, the analysis does not attempt to reconstruct the exact narrative-threat values or dynamic-difficulty parameters experienced by individual participants.

\subsection{User Study Results}

The interviews revealed four recurring themes: the perceptibility of the Rise/Fall signal across narrative and gameplay, the role of runtime adaptation in shaping gameplay experience, the coherence and integration of generated content, and the system's potential as a creative-support and rapid-prototyping tool.

\subsubsection{Rise/Fall provided a recognizable connection between narrative direction and gameplay.}
Participants engaged with Rise/Fall both as a representation of narrative progression and, in several cases, as a signal of gameplay tension. P4 explicitly observed that ``the Rise levels were definitely easier than the Fall levels,'' while P12 associated Fall with greater tension and harder combat, although they were uncertain whether the difference was attributable specifically to Rise/Fall. P7 similarly described a Fall level that led directly to a boss fight while the corresponding Rise level contained regular enemies. P10 provided a more detailed example in an ocean-themed game: the Rise level took place on an open tidal flat with a wider field of view and fewer enemies, while the Fall level descended into the dark deep sea with more monsters and greater pressure. P10 felt that this tension ``matched the ocean story very well.'' These accounts indicate that the intended mapping from narrative trajectory to gameplay could remain perceptible after being propagated through the generation pipeline.

Participants also used Rise/Fall to reason about narrative direction. After recognizing that doors corresponded to narrative nodes, P16 began choosing according to the desired story trajectory rather than combat difficulty, explaining that ``In the node view, the upward and downward structure is quite clear, and you can visually see the development of the story.'' P15 similarly became more attentive to Rise/Fall after understanding its meaning and began anticipating greater danger when continuing downward through Fall nodes. This suggests that the abstraction supported not only system-side orchestration but also player interpretation and decision-making.

The strength of this perception nevertheless varied across participants. P1 reported not noticing the distinction between Rise and Fall levels and felt that the levels were approximately the same difficulty. P6 similarly described Rise and Fall as ``about the same.'' P14 recognized the different Rise/Fall doors but felt that their gameplay consequences could have been communicated more explicitly, explaining that the choice did not feel sufficiently significant and was therefore made casually. P16 also distinguished between the two stages of the system, finding the Rise/Fall representation clear in the node view but concluding that ``it isn't expressed strongly in the gameplay, but for creation it is useful.'' Taken together, the interviews suggest that Rise/Fall remained recognizable across the generated experience, with particularly strong legibility as a representation of narrative direction and positive, though less uniform, evidence of its intended gameplay consequences.

\subsubsection{Runtime adaptation provided subtle rather than explicit gameplay modulation.}
Participants' accounts suggest that runtime difficulty adaptation operated primarily as a subtle layer of gameplay modulation rather than an explicitly identifiable mechanic. P10 provided a particularly clear account, describing enemy pressure increasing when they were performing well and easing when health was low. P10 explained that this adjustment made the game feel smoother and reduced the likelihood of becoming stuck, closely matching the intended adaptive behavior. P4 also reported noticing that level difficulty or enemy attacks adjusted according to performance.

Other participants did not explicitly attribute changes in difficulty to adaptation. P12 noted that unfamiliarity with the combat system made such changes difficult to judge, explaining that they had not yet understood the combat system well enough to determine whether difficulty was adapting to their performance. P14 did not report noticing adaptation, and P16 similarly did not identify enemy attacks as changing according to performance. Importantly, these responses distinguish conscious recognition of adaptation from its potential experiential effect: because adaptation is integrated with encounter variation, Rise/Fall differences, and differences in player performance, participants were not necessarily able to isolate its contribution during play. The results therefore suggest that runtime adaptation can contribute to responsive gameplay while remaining largely unobtrusive, while also indicating an opportunity to make adaptive consequences more legible when explicit player awareness is desired.

\subsubsection{Generated content successfully expanded initial ideas into coherent playable experiences.}
Participants were broadly positive about the system's ability to transform limited initial input into substantially developed narrative and visual content. P12 was impressed that a simple prompt could produce an entire story and evaluated the generated characters, objects, plot, dialogue, and visuals positively. P16 similarly observed that the system developed their initial premise further than expected, remarking during generation that the system had developed the idea further than they had and introduced narrative developments they had not anticipated. These responses support the system's ability to expand sparse authorial input into a richer structured narrative and instantiate it as a playable experience.

Participants also identified opportunities to improve the transition from generated narrative content to spatial gameplay. P14 found narrative information during play relatively fragmented. Two participants (P5 and P8) identified specific cases in which narrative descriptions, transitions, and gameplay behavior could have been more closely aligned. P5 encountered a character described as the protagonist's companion, but the character did not subsequently follow the protagonist during gameplay, which P5 found confusing. P8 identified a different transition problem: one scene presented a king choosing between two options, while the following scene moved directly to a conversation with another character without clearly connecting the two events. These observations point primarily to the challenge of presenting and spatially instantiating generated content rather than to the generation of the underlying content itself.

A similar distinction appeared in participants' responses to generated visual assets. Participants generally accepted the generated assets, but five participants (P1, P6, P12, P13, and P16) identified problems related to high object density or environmental placement. P1 reported that forests, rocks, and other environmental elements could obscure enemies and make their attacks difficult to see. P6 found that dense plants and other objects made paths unclear. P12 observed that there could be ``so many decorations on the scene'' that both the player and boss became difficult to see. P13 similarly reported that environmental elements could obscure enemies and make enemies difficult to distinguish from NPCs. P16 described excessive object density as ``a cluttered mess'' and argued that navigable paths should remain visually clear. These responses suggest that the system can generate sufficiently rich visual content for playable instantiation, while future orchestration mechanisms could further optimize spatial composition according to gameplay constraints such as visibility, navigation, and combat readability.

\subsubsection{The system showed strong potential for ideation and rapid prototyping.}
Participants repeatedly identified creative exploration and rapid prototyping as compelling uses of the system. P1 described it as a potential ``rapid validation or rapid prototyping tool,'' and proposed using the system to first create a lightweight 2D version of a game concept to test its gameplay and numerical design before moving toward a more expensive 3D implementation. P8 independently described the system as particularly suitable for taking a minimal game idea and immediately turning it into a playable prototype. Rather than requiring authors to specify a complete game structure, the system allowed an initial premise to be expanded into alternative narrative trajectories and instantiated into playable content.

Participants also identified ways in which greater authorial control could strengthen this workflow. P15 proposed adding or deleting nodes, constraining unwanted narrative directions, and controlling how strongly generation followed the initial prompt. P16 similarly emphasized the value of editable narrative nodes and suggested that the structured planning process could extend beyond games to novels and other narrative forms. P16 specifically considered the Rise/Fall representation useful for creation, noting that it ``isn't expressed strongly in the gameplay, but for creation it is useful.''

These responses position the system not only as an automated generation pipeline, but as a potential mixed-initiative creative environment in which generation exposes possibilities that authors can subsequently inspect, select, and refine. Participants' requests for additional control therefore suggest a natural extension of the current approach toward more interactive forms of whole-game generation.

Overall, the interviews provide evidence that the shared semantic signal remained meaningful beyond the generation pipeline. Rise/Fall was particularly legible as a narrative-planning abstraction and, for participants including P4, P7, P10, and P12, was also associated with the intended differences in gameplay tension and difficulty. P10's experience further demonstrated how narrative direction, environmental presentation, encounter pressure, and player choice could align around the same Rise/Fall signal. Runtime adaptation was less consciously identifiable, but P10 independently described behavior closely matching its intended operation, suggesting that adaptation can function as a subtle layer of gameplay modulation. More broadly, participants responded positively to the system's ability to expand initial ideas into structured narratives and playable experiences, and P1 and P8 identified rapid prototyping as a concrete application. The remaining challenges primarily concern the legibility, spatial integration, and authorial control of generated content, providing clear directions for extending the orchestration framework.

\section{Discussion}

\subsection{Propagation of Narrative Archetypes}

Our results suggest that narrative archetypes can serve as persistent semantic signals for coordinating heterogeneous stages of complete game generation. In Forking Garden, the Rise/Fall representation is first introduced during narrative planning and subsequently influences path selection, gameplay generation, and runtime adaptation. Our technical evaluation shows that this signal produces measurable downstream effects at the level of game mechanics, including an association between narrative threat and encounter danger. The user study further supports this result: multiple participants independently associated Rise/Fall with differences in gameplay tension or difficulty, suggesting that the effects of the signal remain observable even after it has undergone multiple transformations throughout the generation pipeline.

The significance of this result lies less in the specific Rise/Fall distinction than in the use of a compact narrative abstraction as an intermediate representation. Complete game generation requires narrative, encounters, objectives, rewards, and runtime systems to make mutually compatible decisions despite operating with different representations. Rather than requiring every component to reason over the complete narrative, a narrative archetype provides a shared semantic state that downstream components can interpret according to their respective generation tasks. At the same time, a shared symbolic representation preserves the entities and gameplay decisions produced by these stages during content instantiation. This separation allows semantic coordination to persist without requiring every generated asset or implementation detail to directly encode the narrative archetype.

The user study also suggests that this propagated signal can acquire meaning at multiple levels of the final experience. Some participants directly associated Rise/Fall with changes in gameplay tension or difficulty, while others interpreted it primarily through narrative progression and branching structure. This distinction is important because successful semantic propagation does not require every downstream manifestation to be interpreted by players in exactly the same way. Instead, the same shared signal can support consistency across different generation stages while allowing players and creators to interpret the resulting experience in different ways.

\subsection{Player Perception}

Participants perceived the propagated narrative signal at different levels of the generated experience. Some directly connected Rise/Fall to gameplay outcomes. P4 perceived Rise levels as easier than Fall levels, while P12 associated Fall with greater tension and more difficult combat. P10 even used this information strategically when choosing branches, selecting Rise when their health was low and Fall when they were prepared for greater difficulty. These responses suggest that the narrative distinction did not remain confined to the system's internal generation pipeline, but continued to carry meaning during actual interaction.

Other participants primarily noticed differences between encounters, or became more attentive to Rise/Fall only after understanding the branching structure. Importantly, perceiving gameplay variation is not the same as attributing that variation to the narrative archetype. P8 noticed that some levels were easier and others more difficult but did not consciously associate this difference with Rise/Fall while playing. P15 and P16 also reported that their subsequent choices became more deliberate once they understood what the branching structure represented. These observations suggest a progression from \textit{generated difference}, to \textit{perceived difference}, to \textit{semantic attribution}. Evidence of all three appeared in our study, indicating that a propagated narrative signal can influence player experience even when players differ in how explicitly they recognize its semantic source.

Runtime adaptation revealed a similar distinction between mechanical effects and explicit attribution of their source. P10 observed that enemy pressure increased when they were performing well and decreased when their health was low, closely matching the system's intended adaptive behavior. Other participants did not explicitly recognize that gameplay changes resulted from runtime adaptation. Because adaptation occurs alongside differences between encounters, Rise/Fall conditions, and changes in the player's own performance, its effects may be experienced without being isolated as a distinct mechanism. This suggests that adaptive behavior can operate as a relatively implicit layer of the generated experience, while also raising a design question: when should such adaptation remain unobtrusive, and when should its effects be communicated more explicitly to players?

Differences among participants may also reflect how they allocated attention during play. Participants who treated branches as meaningful decisions (for example, by reading narrative descriptions or considering expected difficulty) appeared more likely to connect subsequent gameplay experiences with Rise/Fall. However, our study did not systematically measure participants' gaming experience, narrative preferences, or player types, so we cannot determine which individual factors account for these differences. Future studies could explicitly measure these characteristics to investigate how player traits and information presentation affect the progression from generated difference, to perceived difference, to semantic attribution.

\subsection{Interpretability from a Creator Perspective}

The interviews suggest that narrative archetypes may be particularly suitable as creator-facing abstractions. When Rise/Fall was presented as part of the narrative structure, participants found it highly interpretable and could explicitly reason about the direction in which a generated story was developing. P16, for example, found the rising and falling progression in the node view particularly clear and considered the representation useful for understanding how a story develops. This suggests that narrative archetypes can function not only as internal coordination signals but also as interpretable abstractions through which creators can inspect the overall trajectory of a generated narrative.

Participants' requests for greater creative control further reinforce this role. P15 wanted the ability to add or remove nodes, restrict undesirable narrative directions, and control how closely generated results adhere to the initial prompt. P16 similarly emphasized the possibility of editing narrative nodes and continuing generation after manual intervention. These requests suggest that, rather than treating generation as a one-way transformation from a prompt to a complete game, creators may benefit from interacting directly with this intermediate semantic representation. Narrative archetypes provide a potential level for human--AI collaboration: they are substantially simpler than directly editing implementation-level game content, while still providing meaningful control over narrative direction.

This also helps explain why participants frequently viewed Forking Garden as a tool for ideation or rapid prototyping. P1 noted its potential for ``quick validation or rapid prototyping,'' while P12 responded positively to the system's ability to generate a complete experience from a simple prompt. Participants' desire to edit and further refine generated content also suggests that the value of complete game generation does not necessarily lie solely in automatically producing a finished playable artifact. Such systems can instead help creators quickly externalize an initial idea, inspect alternative trajectories, experience the gameplay consequences of those trajectories, and then revise the underlying structure.

Narrative archetypes are particularly well suited to this role because they occupy an intermediate level between high-level creative intent and low-level generated content. They are abstract enough to coordinate heterogeneous generators, yet intuitive enough to expose meaningful structure to creators. This dual role suggests a broader possibility for complete game generation: intermediate representations need not remain hidden implementation mechanisms, but can instead become interfaces through which creators collaborate with generative systems.

\subsection{Limitations and Future Work}

The current framework has several limitations that also point toward directions for future work. First, although participants perceived the propagated semantics at both narrative and gameplay levels, they varied in how explicitly they attributed these differences to their underlying source. Rise/Fall was particularly easy to understand when presented through narrative structure, whereas its influence on gameplay and runtime adaptation was more implicit for some participants. Future work could directly compare the system's intended narrative trajectory with the trajectory perceived by players over the course of gameplay, examining whether the intended semantic signal is successfully communicated through narrative structure, gameplay mechanics, and audiovisual cues.

Second, the current system instantiates different narratives within a relatively fixed gameplay framework. Participants noted that different generated stories could still result in similar combat experiences. This suggests that semantic propagation could extend beyond encounter difficulty and enemy composition to the structure of interaction itself. Supporting multiple gameplay structures, richer progression systems, and more diverse encounter and objective types could allow different narrative trajectories to produce more substantially differentiated forms of play.

Third, participants identified room for improvement in the spatial integration of generated content. Content-rich generated environments could sometimes become cluttered, suffer from occlusion or asset repetition, or exhibit mismatches between narrative descriptions and actual gameplay behavior. These observations suggest that although symbolic consistency provides an important foundation for complete game generation, it must be complemented by additional spatial and functional constraints. Future content-instantiation systems could explicitly account for visibility, navigation, combat space, object density, and functional affordances while maintaining semantic consistency.

Fourth, the current study did not systematically record participants' gaming experience, narrative preferences, or play styles. Because participants differed in the degree to which they attended to narrative branches and gameplay cues, these individual characteristics may influence how propagated semantics are perceived and interpreted. Larger-scale studies incorporating explicit measures of player characteristics could further examine this relationship and determine which forms of semantic presentation are most effective for different types of players.

Finally, participants' interest in editing narrative nodes points toward mixed-initiative generation as an important extension. The current framework demonstrates how a semantic signal can automatically propagate across multiple generation stages; future systems could allow creators to intervene directly at this intermediate level and selectively regenerate downstream content while preserving unaffected decisions. In this way, a shared semantic representation could serve not only as a mechanism for game orchestration but also as an interface for controllable complete game generation.

\section{Conclusion}

We presented Forking Garden, a complete game generation framework that uses narrative archetypes as persistent semantic signals to coordinate narrative and gameplay generation. Rather than treating narrative, gameplay, and content generation as independent tasks, Forking Garden propagates a shared Rise/Fall representation from narrative planning through gameplay generation and runtime adaptation. A shared symbolic representation further preserves generated entities and gameplay decisions during content instantiation, allowing semantic relationships established during planning to persist into the final playable experience.

Our evaluation shows that this shared signal produces meaningful effects across multiple stages of the generation pipeline. The technical evaluation demonstrates that narrative constraints shape generated narrative trajectories and that narrative threat is reflected in measurable encounter danger. The user study further shows that these propagated distinctions remain meaningful to players: multiple participants independently associated Rise/Fall with expected differences in gameplay tension or difficulty, while others primarily used the representation to understand and select among alternative narrative trajectories. Runtime adaptation was perceived more implicitly, suggesting that generated mechanical differences can influence player experience even when players do not explicitly attribute those effects to their underlying semantic source.

Participants also broadly valued the system's ability to expand an initial idea into a structured narrative and playable experience, particularly for ideation and rapid prototyping. Their interest in inspecting and editing generated narrative structures further highlights the potential of narrative archetypes as creator-facing representations. Because the same abstraction can both coordinate generation and remain intelligible to human creators, it can provide a potential interaction layer between automated generation and creator control.

These results suggest that complete game generation requires more than generating individually coherent game facets. It also requires a mechanism through which semantic relationships can persist as decisions move across different forms of generation. Forking Garden demonstrates how a compact narrative archetype can serve this role, connecting narrative intent to generated structure, gameplay outcomes, runtime behavior, and creator interpretation. More broadly, this work points toward a class of complete game generation systems in which semantic intent is not merely stated once at the beginning of generation and subsequently discarded, but persists throughout the generation process as an interpretable and actionable representation.

\bibliographystyle{ACM-Reference-Format}

\bibliography{references}

\appendix

\section{Schema-Constrained Art Generation Example}
\label{sec:instantiation-example}

We provide the Flower Girl character as a concrete example of the two-stage
schema-constrained art generation process described in Section~4.4.
Table~\ref{tab:flowergirl-spec} shows the entity-level specification used in
the first stage. This specification establishes the character's visual
identity before any executable geometry is generated. The second stage then
takes this specification as fixed input and expands it into vector layers
that can be directly interpreted by the game engine.

\begin{table*}[t]
\centering
\caption{Stage 1 entity-level art specification for the Flower Girl.}
\label{tab:flowergirl-spec}

\begin{tabularx}{\textwidth}{
    >{\raggedright\arraybackslash}p{0.18\textwidth}
    >{\raggedright\arraybackslash}X
}
\toprule
Schema Field & Flower Girl Specification \\
\midrule

\texttt{silhouette}
& \texttt{humanoid} \\

\texttt{size\_scale}
& \texttt{0.95} \\

\texttt{palette}
& \texttt{primary: \#5CA968},
  \texttt{secondary: \#3E7A46},
  \texttt{accent: \#F2A8C8},
  \texttt{outline: \#2E5A36},
  \texttt{glow: \#FFE9A8} \\

\texttt{features}
& \texttt{twin\_tails},
  \texttt{flower\_hairpiece},
  \texttt{flower\_basket\_prop},
  \texttt{orbiting\_petals},
  \texttt{apron\_dress} \\

\texttt{motion\_style}
& \texttt{sway} \\

\texttt{mood}
& \texttt{gentle} \\

\bottomrule
\end{tabularx}
\end{table*}

\begin{table*}[t]
\centering
\caption{Representative Stage 2 vector-layer instantiations for the Flower
Girl. The complete asset is represented as an ordered sequence of executable
primitive layers.}
\label{tab:flowergirl-layers}

\begin{tabularx}{\textwidth}{
    >{\raggedright\arraybackslash}p{0.15\textwidth}
    >{\raggedright\arraybackslash}p{0.12\textwidth}
    >{\raggedright\arraybackslash}p{0.19\textwidth}
    >{\raggedright\arraybackslash}p{0.13\textwidth}
    >{\raggedright\arraybackslash}X
}
\toprule
Feature & Primitive & Color & Animation & Instantiation \\
\midrule

Apron dress
& Polygon
& \texttt{\#3E7A46}, \texttt{\#5CA968}
& \texttt{sway}
& Nested polygon layers form the outer and inner dress, with additional
layers representing the apron and asymmetric fabric detail. \\

Flower basket
& Ellipse / Circle
& \texttt{\#8A6234}, \texttt{\#B98A4E}, accent colors
& --
& Two ellipse layers form the basket body, while additional circle
primitives represent flowers carried inside it. \\

Twin tails
& Polygon
& \texttt{\#6E4A2E}
& \texttt{sway}
& Separate polygon layers represent the left and right twin tails, allowing
them to move independently from the character's body. \\

Flower hairpiece
& Circle
& \texttt{\#F2A8C8}, \texttt{\#F6C6DC}, \texttt{\#FFE9A8}
& --
& Multiple overlapping circle layers construct the petals and center of the
flower ornament. \\

Orbiting petals
& Circle
& Accent / glow palette
& \texttt{orbit}
& Three petal layers share an orbit radius of 24 but use different phase
values, producing asynchronous motion around the character. \\

\bottomrule
\end{tabularx}
\end{table*}

The generated output also records how the requested entity-level features are
realized in the resulting layer structure. For the Flower Girl,
\texttt{apron\_dress} is mapped to the nested dress and apron layers,
\texttt{flower\_basket\_prop} to the basket and flower layers,
\texttt{twin\_tails} to the hair layers, \texttt{flower\_hairpiece} to the
layered flower ornament, and \texttt{orbiting\_petals} to three independently
animated layers. The recurring flower motif is therefore expressed in several
parts of the character while remaining grounded in the same entity-level
palette and specification.

This example illustrates the separation between identity and executable
realization in the two-stage schema. The first stage specifies the visual
properties that should remain stable whenever the entity appears, while the
second stage determines how those properties are instantiated through
engine-readable geometry and animation bindings. Because these components
remain individually addressable, the generated asset can be rendered and
animated directly at runtime without an additional rigging or sprite-extraction
step.

\end{document}